# Excitonic topological order in imbalanced electron-hole bilayers

Rui Wang[1,2], Tigran A. Sedrakyan[3], Baigeng Wang[1,2] ǂ, Lingjie Du[1,2,4]*, Rui-Rui Du [5,6,7] ξ

[1] *School of Physics and National Laboratory of Solid State Microstructures, Nanjing University, Nanjing 210093, China*
[2] *Collaborative Innovation Center for Advanced Microstructures, Nanjing 210093, China*
[3] *Department of Physics, University of Massachusetts Amherst, Amherst, Massachusetts 01003, USA*
[4]*Shishan Laboratory, Suzhou Campus of Nanjing University, Suzhou 215000*
[5] *International Center for Quantum Materials, School of Physics, Peking University, Beijing 100871, China*
[6]*Collaborative Innovation Center of Quantum Matter, Beijing 100871, China*
[7]*CAS Center for Excellence, University of Chinese Academy of Sciences, Beijing 100049, China*
ǂ *bgwang@nju.edu.cn, *ljdu@nju.edu.cn,ξrrd@pku.edu.cn*

**Correlation and frustration play essential roles in physics, giving rise to novel quantum phases [1-6]. A typical frustrated system is correlated bosons on moat bands, which could host topological orders with long-range quantum entanglement [4]. However, the realization of moat-band physics is still challenging. Here, we explore moat-band phenomena in shallowly-inverted InAs/GaSb quantum wells, where we observe an unconventional time-reversal-symmetry breaking excitonic ground state under imbalanced electron and hole densities. We find a large bulk gap exists encompassing a broad range of density imbalance at zero magnetic field (*B*), accompanied by edge channels that resemble helical transport. Under an increasing perpendicular *B*, the bulk gap persists, and an anomalous plateau of Hall signals appears, which demonstrates an evolution from helical-like to chiral-like edge transport with a Hall conductance ~ $e^2/h$ at 35 Tesla. Theoretically, we show that strong frustration from density imbalance leads to a moat band for excitons, resulting in a time-reversal-symmetry breaking excitonic topological order, which explains all our experimental observations. Our work opens up a new direction for research on topological and correlated bosonic systems in solid states beyond the framework of symmetry-protected topological phases, including but not limited to the bosonic fractional quantum Hall effect.**

**Introduction**. In correlated bosonic systems, novel quantum states could emerge [1-3] when there is a strong frustration effect [4-6]. One pertinent example is given by bosons on the "moat band" [7-10], initially proposed in cold atom systems, where the energy minima of the band constitute a degenerate loop in momentum space. The boson-boson interactions then become dominant, which generate emergent fluxes and favor topological orders with anyonic excitations and spontaneous time-reversal symmetry (TRS) breaking [7-10]. Distinct from symmetry-protected topological phases, the emergent physics here is in analogy with the fractional quantum Hall (FQH) effect [5,11]. Regardless of these fascinating prospects, the implementation of moat bands is still challenging in cold atom experiments.

In semiconductors, it is well-known that electrons and holes can form bosonic pairs called excitons. The excitons can spontaneously form in electron-hole bilayers [12], providing a new avenue to realize the moat band physics. In systems with balanced electron-hole densities, Bardeen-Cooper-Schrieffer (BCS)-like condensation of zero-momentum excitons is favored (Fig.1a), resulting in topologically trivial exciton insulators (EIs) [2,3,13]. For imbalanced cases, the excitons formed would carry finite momentum *Q*, giving

finite-$Q$ EIs [14-21]. Interestingly, we find that frustration would be strongly enhanced in this case since one electron (hole) could form pairs with different individual holes (electrons), leading to a large number of competing excitonic configurations with close energies (*i.e.*, frustrations for excitons). As described below, an excitonic moat band displaying energy minimum at $|\mathbf{q}|=Q$ (Fig. 1b) would emerge [22] in favor of TRS-breaking topological orders.

In shallowly-inverted InAs/GaSb quantum wells (QWs) (Fig. 1c) where electron ($n_e$) and hole ($n_p$) densities were tuned by gate voltages (Fig. 1d) to ~ $5.5 \times 10^{10}$ cm$^{-2}$, a BCS-like EI gap was reported by optical and transport measurements [12], suggesting the existence of an EI ground state. Surprisingly, in the EI gap, a pair of counter-propagating channels, *i.e.*, helical-like edges, was discovered in mesoscopic devices [23] with quantized edge transport described by Landuer-Büttiker (LB) formula [24], implying a topological state (phenomenologically termed as topological EI). However, no evidence for the destruction of edge states was found for magnetic fields up to 10 Tesla [23, 25 and Sec. I of Supplementary Information], which seems against the notion within the single-particle picture that topological insulators with helical edge transport should be TRS-protected and may not survive under magnetic fields [26-29]. Notice that, at such low densities, even at the charge-neutral-point (CNP), because of potential fluctuations, electron-hole densities tend to be imbalanced locally [30]. Consequently, frustration and the excitonic moat band can emerge, posing a key question, *i.e.*, whether the topological EI could be generated by the moat band with TRS breaking. Thus, investigating the roles of TRS and density imbalance in the topological EI, which requires experiments on gated devices with high magnetic fields, is highly desired.

Here, we utilize magnetic fields as high as 35 T and a broad range of gate tunability of the exciton momentum to reveal the topological nature of the EI state in shallowly-inverted InAs/GaSb QWs. We find that the EI gap unexpectedly persists from electron-dominated to hole-dominated regimes. Correspondingly, under perpendicular magnetic field $B_\perp$, an anomalous plateau of Hall signals is observed in the same regime. Remarkably, $B_\perp$ dependence of the Hall signals in the plateau presents helical-like to chiral-like evolution of edge transport with Hall conductance close to $e^2/h$ for high $B_\perp$ up to 35 T. The data provide strong evidence for a TRS-breaking EI formed under e-h imbalance beyond the description of existing theories. Theoretically, we show that an excitonic moat band naturally occurs due to the frustration emergent from density imbalance. This brings a TRS-breaking excitonic topological order (ETO), providing a unified explanation for all the observed topological excitonic properties. Our results demonstrate that the unconventional excitonic state could be a topological order intrinsically associated with correlated excitonic states.

**Finite momentum excitonic insulator.** In shallowly-inverted InAs/GaSb QWs (see Methods), the bulk conductance $\sigma_c$ is measured from a Corbino device A where edge states are shunted by contacts. At 0 T, as $n_e$ is depleting by the front-gate voltage $V_f$ to match $\sigma_c$, $\sigma_c$ reaches zero around -2.5 V due to the formation of the topological EI [12] (Fig. 2a). The zero $\sigma_c$ persists over a wide $V_f$ range from ~ -2 V to ~ -3 V, where we perform $T$-dependence experiments. Within this range, $\sigma_c$ exhibits exponentially activated $T$-dependence, allowing extraction of gap energy $\Delta$ which is nearly symmetrical with respect to the sign of the net charge (Fig. 2a and Extended Data Fig. 1); by contrast, outside this range, $\sigma_c$ is finite and weakly dependent on $T$, indicating metallic transport in the two-carriers regime. Noted that at $V_f \sim$ -2 V, $n_e \sim 1.1 \times 10^{11}$ cm$^{-2}$ and $n_p \sim$ $4.5 \times 10^{10}$ cm$^{-2}$ (Extended Data Fig. 2), therefore the e-h densities are strongly imbalanced, favoring finite-$Q$ excitons; at $V_f \sim$ -3 V the imbalance is reversed, $n_p \gg n_e$. At $V_f \sim$ -2.5 V, the averaged e-h densities are approximately equal, $n_e \sim n_p \sim 5.5 \times 10^{10}$ cm$^{-2}$, but as mentioned earlier, the carriers would still be considered as locally imbalanced [30] (see Methods).

At 0 T, $\Delta$ reaches a maximum of ~ 25 K and reduces to ~10 K around the imbalanced points ($V_f$ ~ -2.2 V or -2.8 V), confirming that the bulk gap remains open from the e-dominated to the h-dominated regime. The persistence of the gap suggests that the ground state stays at the same gapped EI phase, regardless of the varying net charge. Applying $B_\perp$, we observe that the gap behavior is essentially the same, but the gap maximum shows a slight increase from 0 T to 2 T and is more visibly enhanced for 4 T and 6 T.

Hall measurements are subsequently performed in a Hall-bar device B (Fig. 2b). For macroscopic devices, at zero field, the helical-like transport still exists with a coherent length $\lambda$=4.4 μm [23], implying transport through a series of quantum resistors, $R_q = h/e^2$ (Extended Data Fig. 3). At fixed $B_\perp$= 4,6,8 T and sweeping $V_f$ from -0.5 V to -2 V, we observe several quantum Hall (QH) plateaus of electrons, albeit they are mixed with signals from minority holes. Using the Onsager relation, we extract the asymmetric term $R_{xy}^{as}$ to obtain the Hall component ($R_{xy} \approx R_{xy}^{as}$ for $B_\perp$>8 T) and the Hall coefficient $R_H = R_{xy}^{as}/B_\perp$ (Fig. 2c) (see Methods). Remarkably, under a fixed $B_\perp$, as $V_f$ tunes carrier densities across the CNP, $R_H$ (and thus $R_{xy}^{as}$) is nearly independent of carrier densities in a large range encompassing e-dominated ($V_f$ ~ -2 V) to h-dominated ($V_f$ ~ -3 V) regimes, i.e., a plateau of Hall signals. Notably, for $V_f$ from -0.5 V to ~ -2 V, $R_H$ suddenly collapses into the plateau, indicating a phase transition occurring at a critical e-h density imbalance. Notice that this Hall plateau already occurs at 2 T (Figs. 2b and 2c), whereas the QH effect has not shown up, suggesting that the plateau is not of QH origin and is anomalous (see Methods). Our results provide strong evidence that the topological EI is a finite-momentum EI from imbalanced electrons and holes across the entire plateau.

**Helical-to-chiral-like edge evolution.** Figure 2d shows the longitudinal resistance peak $R_{xx}^{EI}$ in the topological EI regime ($V_f$ ~ -2 V to ~ -3 V) decreases with $B_\perp$, characterized by an increasing $\lambda$ against $B_\perp$ (~ 50 μm at 18 T, see inset of Fig. 2d). Under $B_\perp$, the Lorentz force imposed on the edge channels would push the one with opposite chirality to $B_\perp$ into the bulk (inner-loop) and leave the other with the same chirality on the boundary (outer-loop). The two separated channels account for suppressed transmission rates between them (*i.e.*, backscattering) and decreased $R_{xx}^{EI}$ (Sec. II of Supplementary Information).

For helical-like edge transport, contributions from channels of opposite chiralities would cancel and give nearly zero Hall signals. Under increasing $B_\perp$, transmission rates between the inner-loop and outer-loop/metal contacts would decrease while those between the outer-loop and metal contacts would remain as unity, which describes lifted Hall signals (inset of Fig. 2c). In the nearly separated case, these two factors lead to chiral-like transport dominated by the outer-loop, characterized by $\sigma_{xy}$~ $e^2/h$. This helical-like to chiral-like edge transport evolution under $B_\perp$ is a hallmark of the topological EI. Here, different from chiral QH edge states, the presence of the inner-loop makes the backscattering between the two channels hardly eliminated so that precise Hall quantization would not be expected under $B_\perp$ (Sec. II of Supplementary Information).

We perform a systematic experiment of $\sigma_c$ vs. $V_f$ vs. $B_\perp$ (0≤$B_\perp$≤35 T) in the Corbino device to map out a phase diagram containing the topological EI and the integer QH states. Figure 3a presents a fan diagram where the QH states with zero-conductance have occupied the space (light blue) outside the topological EI regime (light green). Notably, the bulk gap (characterized by zero-conductance) of the topological EI persists continuously up to an extremely high $B_\perp$. As shown in Extended Data Fig.4, the Hall conductance $\sigma_{xy}$ of a Hall-bar device C in the anomalous plateau is lifted from zero for raised $B_\perp$; at high $B_\perp$, i.e.,18 T, $\sigma_{xy}$ becomes close to $e^2/h$. Figures 3b and 3c highlight $\sigma_{xx}$ and $\sigma_{xy}$ measured in a Hall-bar device D, together with $\sigma_c$. Remarkably, in the topological EI regime, $\sigma_{xy}$ shows a plateau around $e^2/h$ at $B_\perp$=16 T and becomes even closer to the quantized value at $B_\perp$=35 T. Since the bulk gap is continuously open from low to high $B$

⊥, the nearly quantized Hall signals at 35 T are evolved from the anomalous plateau observed at low $B_⊥$. In contrast with the low $B_⊥$ case (Fig. 2b), here the Hall signals in the plateau saturate from 16 T to 35 T, ~ $e^2/h$, providing concrete evidence for the emergence of chiral-like edge transport at strong $B_⊥$ in the topological EI regime. The observed chiral-like edge transport clearly demonstrates that the topological state is robust against broken TRS.

The chiral-like transport in the topological EI regime has a distinct nature from that of the $v$=1 QH state (see Methods). In sweeping $V_f$, $\sigma_c$ exhibits a distinct peak separating the topological EI and the $v$=1 QH (white areas in Figs. 3b and 3c) states, clearly signaling opening-closing-reopening of the bulk gap, hence a topological transition [26, 27] occurs. This is also evidenced by a $\sigma_{xy}$ bump and finite $\sigma_{xx}$ in the transition regime, where the gap is closed with two-carrier transport dominating in bulk (see Sec. III of Supplementary Information). Our results reveal an unconventional excitonic ground state formed with TRS breaking under density imbalance, distinct from all topological and excitonic phases studied to date [26-28,31-37] (Sec. VII of Supplementary Information).

**Emergent excitonic topological order.** To understand the nature of the EI, we study correlated electron-hole bilayers with density imbalance [14-21]. Under imbalance, the excess electrons/holes have nontrivial effects on e-h pairs; they intend to break the e-h pairs and form new pairs, leading to a large number of excitonic configurations with competing energies (Fig. 1b). As schematically illustrated by Fig. 4a, in a small system consisting of three electrons and a hole, there occur different pairing configurations with competing energies, in analogy with the frustrated quantum spins on a triangular which exhibit degenerate spin states. Since frustration is known to suppress long-range orders [6], disordered quantum states may also be favored in the excitonic systems.

In the momentum space, two concentric Fermi surfaces emerge due to the imbalance (Fig. 4b). Consequently, formed excitons possess finite momenta on a loop, *i.e.*, |**q**|=Q. The excitons have an equal tendency to condense on the loop [15], and thus have infinite condensation channels. The interaction between excitons and excess particles significantly renormalizes the excitonic dispersion (Sec. IV.7.5 of Supplementary Information), resulting in a moat band as shown by Fig. 1b. Moreover, an exciton-exciton interaction also emerges with the strength $U$, leading to the low-energy effective theory:

$$H_\text{b} = \sum_r b_r^\dagger \left[\frac{(|\hat{q}|-Q)^2}{2m_\text{b}} - \mu_\text{b}\right] b_r + U \sum_r n_{\text{b},r} n_{\text{b},r}, \qquad (1)$$

where $\hat{q} = -i\nabla_r$, $m_\text{b}$ and $\mu_\text{b}$ are the mass, and chemical potential of excitons, $n_{\text{b},r} = b_r^\dagger b_r$ denotes the particle number of excitons. Note that the electrons and holes form weakly-paired BCS states, and large binding energies are not required. Besides, a damping term is implicit in $\mu_\text{b}$, which describes the finite lifetime of the bosons due to their coupling with the excess particles (see Sec. IV of Supplementary Information). Clearly, the emergent bosons are strongly correlated in nature because of the flat dispersion along the moat |**q**|=Q.

The bosonic e-h pairs can be represented as spinless fermions attached to one flux quanta [38,39] (inset of Fig. 4c). The emergent flux behaves as an effective magnetic field $B_\text{CS}$ that generates Landau levels (LLs) (Fig. 4c). This offers a good wave function ansatz for the ground state of $H_\text{b}$, *i.e.*, the lowest LL (LLL) fully filled by composite fermions (CFs). This many-body state is a gapped TRS-breaking topological order arising from the frustrated excitons, dubbed the ETO. Using this ansatz, we find the energy per particle [10]: $E_\text{ETO} = (\pi^2 n_\text{b}^2/2m_\text{b}Q^2)\log^2(4n_\text{b}/Q^2)$ (Sec. V of Supplementary Information). $E_\text{ETO}$ is lower than the energy of the Fulde-Ferrell-Larkin-Ovchinnikov (FFLO) [8] and the non-uniform condensation state [40] for relatively low $n_\text{b}$, which finally leads to the phase diagram in Fig. 4d. Compared to the results at mean-field level (Fig. 4e), the ETO is found to emerge in the intermediate regime between the FFLO and the two-carrier region, as a result of the frustration effect.

The ETO possesses a pair of chiral electron and hole edge modes (Extended Data Fig. 5). Moreover, analogous to BCS superconductors, the particle-hole symmetry would emerge in low-energy around the Fermi level. Under the particle-hole symmetry, the hole channel of the ETO edge can be equivalently cast into an electron mode propagating in the opposite direction, leading to characteristic helical-like transport under zero $B_\perp$. On the other hand, the robustness of ETO under strong $B_\perp$ can be deduced. At the microscopic level, both electrons and holes will occupy the degenerate states characterized by the orbital angular momentum (OAM) in their respective LLL. The electron-hole interaction projected to the LLL then drives the pairing. Interestingly, similar to the moat band physics under $B_\perp = 0$ T, we find a strongly-frustrated regime taking place in the OAM space (Sec. VI of Supplementary Information), which implies the stability of the ETO under high fields, consistent with the persisting EI regime under $B_\perp$ in Fig. 3a.

The ETO formed under a large density imbalance well accounts for the observed gapped EI state, which persists in the wide $V_f$ range from the hole-dominated to electron-dominated regimes. The helical-like edge transport of ETO under zero $B_\perp$ explains the quantized resistance in mesoscopic samples. Importantly, in contrast with topological insulators, the ETO, analogous to the FQH states, has an intrinsic topology that does not rely on any symmetries. Therefore, its edge state is robust against external field $B_\perp$. Intrinsically, the two edge channels of the ETO have opposite chiralities, and $B_\perp$ would drive them spatially separated. Using the LB formula, the calculated resistance of the ETO edge states quantitatively agrees with the data in Fig.2b, 3b, and 3c, well accounting for the helical-like to chiral-like edge transport evolution under rising $B_\perp$ (Extended Data Fig. 6a and Methods). These results provide evidence that the experimentally observed topological excitonic state could be an intrinsic topological order with long-range quantum entanglement. The ETO theory also predicts the emergence of fractional bulk excitations with semionic statistics (Extended Data Fig. 7), which could be explored by experiments beyond magneto-electrical transport, such as heat flow or interferometer measurements. Research along this direction could further demonstrate the emergent moat band and topological orders in other correlated systems.

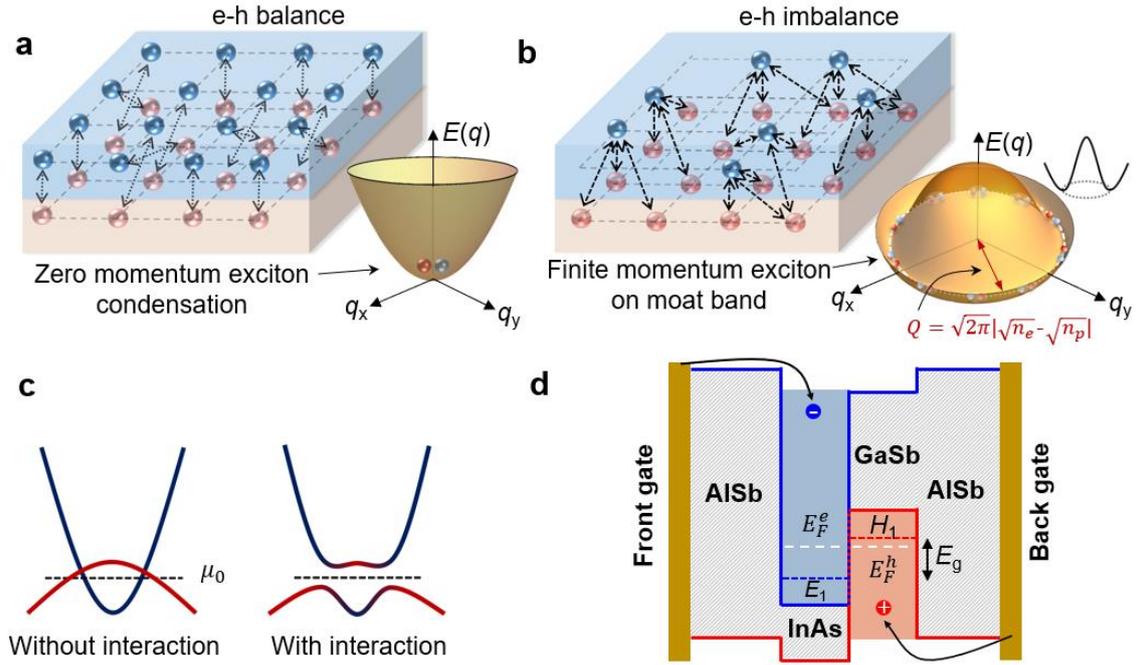

Fig.1 **Finite-momentum excitons in InAs/GaSb quantum wells.** (a) A schematic of a balanced bilayer where $n_e = n_p$, and electrons and holes form BCS-paired excitons with zero-momentum. At low temperatures, excitons can condense into an excitonic insulator. (b) A schematic of an imbalanced bilayer where $n_e < n_p$. An electron could form pairs with different individual holes, leading to a large number of competing configurations with close energies, *i.e.*, frustrations for excitons. A moat band with the energy minimum at $|\mathbf{q}|=Q$ would emerge in the dispersion of excitons. $Q$ is determined by the carrier densities in experiments, $Q=\sqrt{2\pi}|\sqrt{n_e} - \sqrt{n_p}|$. (c) Schematic of the shallowly-inverted InAs/GaSb QW band structure with and without Coulomb interaction. (d) Structure of InAs/GaSb quantum wells which has a broken-gap band alignment, with $E_g = H_1 - E_1 > 0$. Electron and hole densities are controlled by gates. In our experiments here, the back gate voltage is set to zero [12].

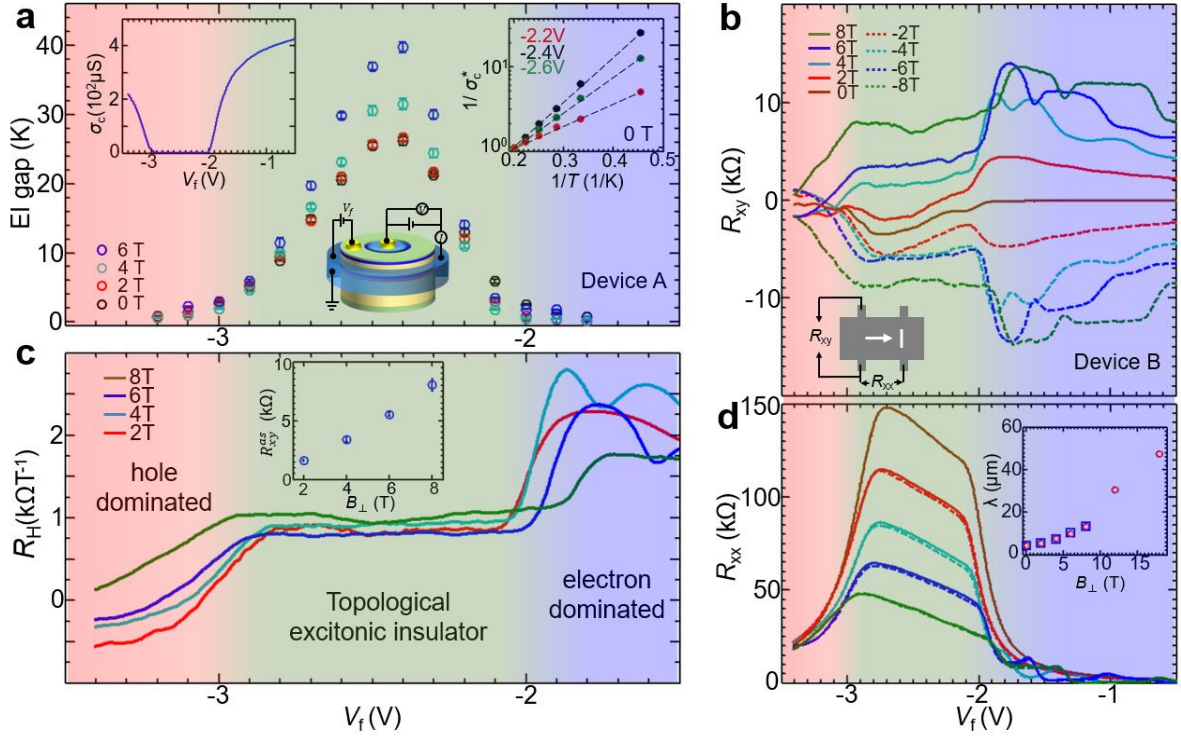

Fig. 2. **Gap energies and anomalous Hall plateau in topological-EI.** (a). EI gap energy $\Delta$ vs. $V_f$ in a Corbino device A. Inset (left) shows $V_f$ dependence of conductance $\sigma_c$ at 0 T and 300 mK. In inset (right), $\sigma_c^*$ is the normalized value by the one at 5 K under 0 T. $\Delta$ is deduced through $\sigma_c \propto \exp(-\Delta/2k_bT)$. More data are shown in Extended Data Fig. 1. Dashed lines are guides to the eye. The results at 2, 4, and 6 T follow a similar process. The cartoon inset sketches the measurement setup. The error bars come from the uncertainty in the extraction of $\Delta$ from the Arrhenius plot. The electron-dominated region is marked by light blue, the topological EI regime by light green, and the hole-dominated region by light pink. (b). Hall resistance $R_{xy}$ vs. $V_f$ in a 25 μm × 50 μm Hall bar device B at 300 mK under $B_\perp$. In the topological EI regime, residual $R_{xy}$ at 0 T is due to mixing with $R_{xx}$ from a slight asymmetry of the Hall-bar (see Sec. II of Supplementary Information). (c). Hall coefficient $R_H$ vs. $V_f$. Inset shows $B_\perp$ dependence of $R_{xy}^{as}$ in the topological EI regime. The error bars come from the uncertainty of $R_{xy}^{as}$. (d). Longitudinal resistance $R_{xx}$ vs. $V_f$ in device B at 300 mK under $B_\perp$. In the topological EI regime, since the bulk is insulating, the $R_{xx}$ is from the edge transport which decreases under increasing $B_\perp$, consistent with the helical-like to chiral-like edge transport evolution. Inset displays the edge coherent length $\lambda$ based on the $R_{xx}$ peaks. In the inset, the blue squares are from data in (d), and the red circles are from data in Extended Data Fig. 4.

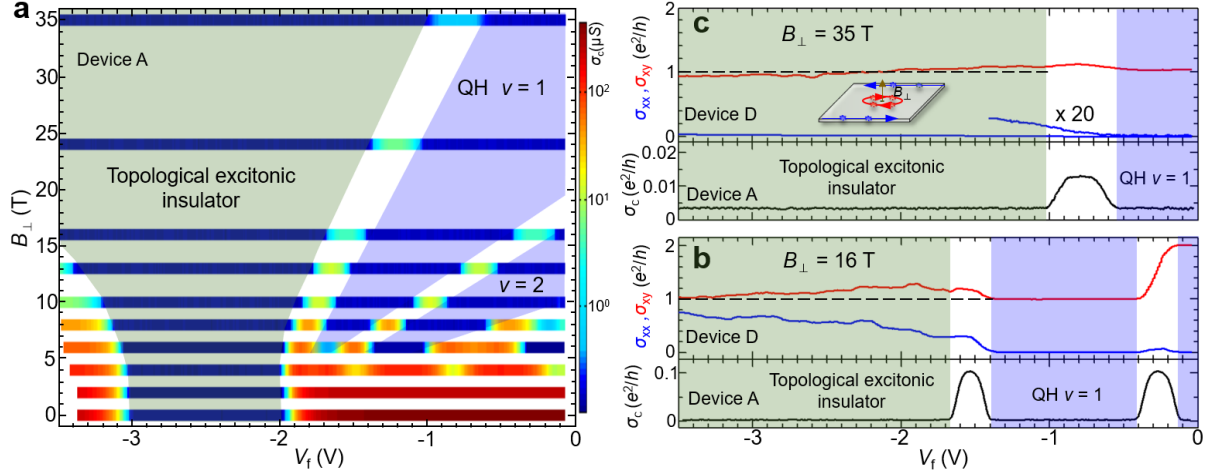

Fig. 3. **Transport under high magnetic fields.** (a). Front-gate voltage $V_f$ dependence of conductance $\sigma_c$ measured in the Corbino device as a function of $B_\perp$ at 30 mK. The color code is exponential with $\sigma_c$. (b) and (c). $\sigma_{xx}$ (blue curves) and $\sigma_{xy}$ (red curves) in a 50 μm × 100 μm Hall-bar device D and $\sigma_c$ in the Corbino device (black curves), taken at 30 mK under 16 T and 35 T, respectively, where wide ranges of zero-conductance can be seen punctuated by peaks. At 16 T, the $\sigma_c$ peak at $V_f$ = -0.25 V separates the $v$ = 2 and $v$ = 1 QH states, and the $\sigma_c$ peak at $V_f$ = -1.55 V separates the $v$ = 1 QH state and the topological EI. Both peaks have the same shape. Corresponding signals in $\sigma_{xx}$ can also be seen, albeit the $\sigma_{xx}$ peak at $V_f$ = -1.55 V is smeared due to the residual conductance in the topological EI regime (Extended Data Fig. 4). Together with the $\sigma_c$ peak at $V_f$ = -0.8 V for 35 T, the data clearly demonstrate a topological transition between the QH state and topological EI. Finite $\sigma_{xx}$ in the transition regime agrees with the values of $\sigma_c$, confirming the gap is closed. The $\sigma_{xy}$ bump in the transition regime is small because the $\sigma_{xy}$ values in the QH and the topological EI regimes are close. The small offset of $\sigma_c$ away from zero is from the detection limit of lock-in equipment. The Inset of (c) depicts separated edge channels under $B_\perp$ due to their opposite chiralities. The light green marks the topological-EI regime, the light blue marks the QH regime, and the white marks the topological phase transitions. In (c), a portion of $\sigma_{xx}$ is multiplied by 20 for clarity.

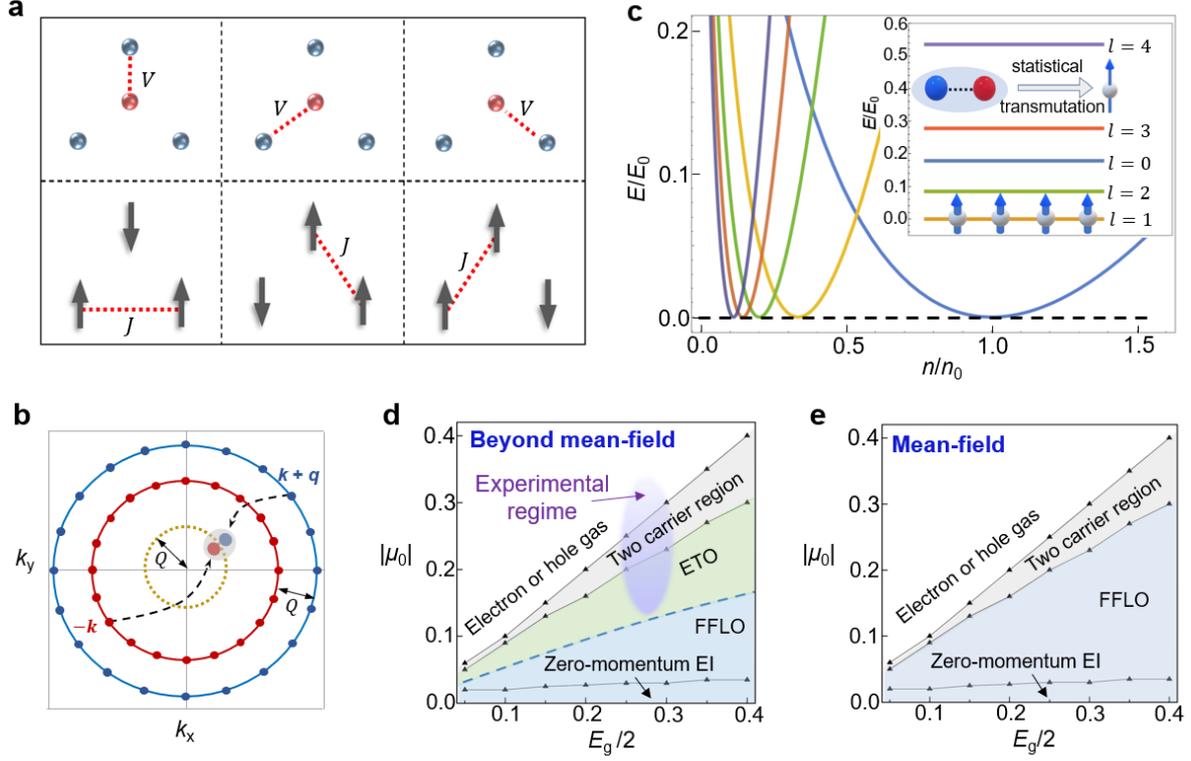

Fig. 4. **The mechanism for excitonic moat band and ETO.** (a) The schematic plot shows the conceptual analogy between the frustrated spins on a triangle and the excitonic system with imbalanced electrons/holes. (b) The excitons tend to condense with equal weight on a momentum loop (dashed circle) as a result of the two concentric Fermi surfaces. (c) Under strong frustration, the e-h pairs can be represented by spinless fermions attached to one flux quanta. The flux then generates Landau levels of the fermions. Taking $n = n_0/3$ as an example, where $n_0 = Q^2/2\pi$ and $E_0 = Q^2/2m_b$, the corresponding ground state is described by the fermions that fully fill the $l = 1$ Landau level. (d) The phase diagram of the original e-h model (see Methods) beyond the mean-field level with varying the band offset $E_g$ (Fig. 1d) and the chemical potential $\mu_0$ that determines the e-h density imbalance. The ETO phase emerges in between the FFLO and the two-carrier region after considering the frustration effect. The shaded purple area denotes the estimated parameter regime accessed by the experiment. Note that the potential fluctuation induces local density imbalance $\sim 1\times10^{10}$ cm$^{-2}$ even at the CNP, which obscures observation of the FFLO and zero-momentum EI phase. (e) The calculated phase diagram of the correlated bilayer model at the mean-field level.

## Methods

**Sample structure and transport measurements.** Our devices were fabricated from shallowly-inverted InAs/GaSb QWs [12,23] grown by molecular beam epitaxy with Corbino (device A) or Hall-bar (devices B, C, D) patterns (Sec. III of Supplementary Information). Here a special metal-contact-buried device architecture (Supplementary Fig. S7) was processed on devices A and D to ensure reliable measurements under an extremely high $B_\perp$. Since devices A and D were fabricated together, they had the same gating parameters and thus can be directly compared. The InAs/GaSb wafer was prepared by molecular beam epitaxy. The wafer has the following structure: N + GaAs (001) substrate, 1 μm buffer layer, and 10 nm GaSb/12.5 nm InAs QWs sandwiched between 50 nm $Al_{0.8}Ga_{0.2}Sb$ barriers; the interface between the GaSb and InAs QWs was doped with a dilute sheet of Si with a concentration of $\sim 1 \times 10^{11}$ cm$^{-2}$. The systematic data reported here were measured from four devices, including 1 Corbino disc and 3 Hall bars. Electrical transport measurements were mainly performed in the National High Magnetic Field Lab (NHMFL), utilizing a dilution refrigerator (base temperature 20 mK) fitted with an 18 T superconducting solenoid, a dilution refrigerator (30 mK) fitted with a 35 T Bitter magnet, and a He$^3$ refrigerator. In order to improve the signal-to-noise ratio, a low-frequency lock-in technique with an alternating 17 Hz, 20 nA to 100 nA current was employed.

**Extraction of Hall components.** Using the Onsager relation, we extract the asymmetric term $R_{xy}^{as}(B_\perp) = \frac{R_{xy}(B_\perp) - R_{xy}(-B_\perp)}{2}$ and symmetric term $R_{xy}^{s}(B_\perp) = \frac{R_{xy}(B_\perp) + R_{xy}(-B_\perp)}{2}$, where $R_{xy}^{as}$ represents the Hall component and $R_{xy}^{s}$ is from the mixing of $R_{xx}$. Figure 2c shows the Hall coefficient the $R_H \equiv \frac{\Delta R_{xy}^{as}(B_\perp)}{\Delta B_\perp} = \frac{R_{xy}^{as}(B_\perp)}{B_\perp}$ as a function of $V_f$. As shown in Extended Data Fig. 8, at 2 T, $R_{xy}^{s}$ in the plateau, which is determined by $R_{xx}^{EI}$, has similar values with that at 0 T, suggesting that the transport can still be characterized as helical-like; for 2 T<$B_\perp$<8 T, $R_{xy}^{s}$ is quickly suppressed, consistent with decreased $R_{xx}^{EI}$; at 8 T, $R_{xy}^{s}$ is overwhelmed by increased $R_{xy}^{as}$, i.e., $R_{xy} \approx R_{xy}^{as}$.

**Characterization of devices and determination of electron-hole imbalance.** Devices A and D are fabricated with a metal-contact-buried architecture (supplementary Fig. S7) to ensure reliable measurements under an extremely high $B_\perp$. The devices were fabricated together and had the same parameters in gating. In this architecture, metal contacts are extended to overlap with a metal gate; the contacts and gate are separated by an insulating layer of $Si_3N_4$. The $Si_3N_4$ layer was grown by plasma-enhanced chemical vapor deposition. Since the overlap area is not small (several μm$^2$), a high-quality insulating layer is required to avoid shorting between gates and contacts. The architecture is critical for our experiments because those connections otherwise made of two-dimensional electron (hole) gas, as commonly used in Hall bar devices, might accidentally become highly resistive under extremely high $B_\perp$. It is critical that for contacts/arms made of the alloy metal, such failure could be suppressed. More details of device fabrication can be found in Sec. III of Supplementary Information. Characterizations of the wafer can be found in Ref. 23.

Extended Data Fig. 2a shows $B_\perp/eR_{xy}$ vs. $V_f$ for an asymmetric Hall device made from the same wafer. As the top of the hole band is reached by the Fermi level, holes are introduced. Due to the small effective mass of electrons, electrons cannot thoroughly screen the gate field, and the gate would also modulate the hole density. As two-carrier transport dominates, $B_\perp/eR_{xy}$ trace would diverge, and the gate tunability to electrons (holes) is decreased (increased). From the data, we deduce that at the CNP, the densities are ~ 5.5

×10$^{10}$ cm$^{-2}$. For the e-h imbalanced regime across the CNP, a plateau-like feature is observed, indicating the formation of an EI gap. As electrons are nearly depleted, the front gate would mostly tune the holes. We find that, as commonly observed in these devices [41], it is difficult to deplete carrier density by gate voltage below a critical density $n_e^c$ or $n_p^c$, which can be explained by the fact that at the critical density, local potential fluctuations would localize carriers in the potential valley, and percolation transport will take over. For hole carriers in the GaSb layer, we find $n_p^c$ is around 1~2×10$^{10}$ cm$^{-2}$ [41]. For electron carriers in InAs, $n_e^c$ is smaller because of a smaller density of states (DOS) (DOS=$m/\pi\hbar^2$, where $m=m_e$~0.032$m_0$ for electrons, $m=m_p$~0.136$m_0$ for holes, $m_0$ is the bare electron mass). We arrive at an order of magnitude estimation ±1×10$^{10}$ cm$^{-2}$ for the local density imbalance around CNP, and we can estimate the potential fluctuations as ~1.5 K according to the DOS. The local density imbalance of ~1.5 K can be used to characterize the local density inhomogeneity of the sample, which is negligible compared to the gap energies. Besides, the local density imbalance (i.e., fluctuations of the local Fermi level) around CNP has also been revealed by Coulomb drag experiments [30].

In order to determine the electron ($n_e$) and the hole ($n_p$) densities as a function of $V_f$, $R_{xy}$ at 1 T and $R_{xx}$ at 0 T are measured in an asymmetric Hall device, and the data are fit by a semi-classic model of two-carrier transport. Based on the model, $R_{xy} = \frac{B_\perp[(n_p - n_e b^2) + \mu_e^2 B_\perp^2 (n_p - n_e)]}{e[(bn_e + n_p)^2 + \mu_e^2 B_\perp^2 (n_p - n_e)^2]}$, $R_{xx} = \frac{1}{e(n_e \mu_e + n_p \mu_p)} \frac{L}{W}$, where $b=\mu_e/\mu_p$, $\mu_e$ ($\mu_p$) is the mobility of electrons (holes), $L$($W$) is the length(width) of the carrier area. Here $L=W$, $\mu_p$~10$^3$ cm$^2$/Vs is taken for hole densities in the order of 10$^{10}$ cm$^{-2}$. From $R_{xy}$ and $R_{xx}$, we can obtain the electron and hole densities through the two-carrier model, as shown in Extended Data Fig. 2b. The density results from the two methods (*i.e.*, Extended Data Figs. 2a and 2b) are consistent, confirming a transition from electron-dominated regime to hole-dominated regime.

**Comparison of the anomalous Hall signals and those in single-particle Hall effect.** We contrast the anomalous Hall plateau originating from the EI state, as shown here, with those expected from single-particle transport. In a single-particle system containing two types of carriers, the Hall resistance is a balance between electron- and hole-signals, therefore the Hall resistance should sensitively depend on $n_e$ and $n_p$. Hence, in the imbalanced regime, the signal should be dependent on the net charge carrier densities. Moreover, the Hall signal from -2 V to -3 V should change its sign. However, what we have observed is a wide plateau with a non-quantized value, suggesting the plateau is not from the single-particle Hall effect. Importantly, the observed density independence of $R_H$ clearly indicates that nearly all the imbalanced electrons and holes originally near the Fermi surface take part in the EI ground state within the plateau regime. Even if there might be individual carriers that do not participate in the EI, the amount of these carriers must be very small and overwhelmed by the EI state. Otherwise, $R_H$ would vary significantly around the CNP.

Under high $B_\perp$, the anomalous Hall plateau manifests chiral-like transport in the topological EI regime, which clearly has a distinct nature from that of the $v$=1 QH state. A QH state can be identified with an integer filling factor, $v= n_e h/eB_\perp$ = 1,2, ... (see Fig. 3a, where $v$ of electrons are marked) whereas the topological EI encompasses low filling factor $v$<<1 at 35 T. While both the topological EI- (light green in Fig. 3a) and QH- (light blue) regimes have shown zero $\sigma_c$ and ~ zero $\sigma_{xx}$, confirming the bulk gap, it is important to note that the QH state is controlled by filling factor $v$ of underlying electrons or holes, however, the topological EI is controlled by $Q$, the momentum of the low-energy excitons. Since a large bulk gap is already opened even at zero magnetic field and remains open continuously under $B_\perp$, the origin of the EI gap cannot be intrinsically attributed to Landau levels as in the case of QH states.

Focusing on the 35 T data (Fig. 3c), we further notice that, while the $v$=1 QH state exhibits a precise

$\sigma_{xy}$ quantization, the respective plateau in the topological EI regime is close but not exactly equal to $e^2/h$, and $\sigma_{xx}$ becomes close to zero but still has residual values. These edge behaviors of $\sigma_{xy}$ and $\sigma_{xx}$ can be attributed to the remaining backscattering in the two nearly-separated channels under high $B_\perp$ (Sec. II of Supplementary Information). In contrast, in the Corbino device, such edge contribution is absent, giving zero $\sigma_c$ in the topological EI regime. Overall, the above observations confirm that the Hall plateau $\sim e^2/h$ is indeed unrelated to a QH state.

**Model Hamiltonian.** The Hamiltonian is given by $H_{eh}=H_0+H_I$. In the basis $\psi_{k,\sigma} = [c_{+,k,\sigma}, c_{-,k,\sigma}]^T$ where $c_{+,k,\sigma}(c_{-,k,\sigma})$ denotes the annihilation operator for the conduction (valence) band electrons with spin $\sigma$, the Hamiltonian $H_0$ depicting the electrons and holes reads as $H_0 = \sum_{k,\sigma} \psi_{k,\sigma}^\dagger (\xi_k \sigma^z - \mu_0)\psi_{k,\sigma}$. $k$ is the wave vector, and the chemical potential $\mu_0$ determines the e-h density imbalance. The energy dispersion is given by $\xi_k=\varepsilon_k-E_g/2$, where $\varepsilon_k=k^2/2m$ with $m$ being the effective electron/hole mass and $E_g$ the energy offset at $k=0$ (Fig. 1d). For the imbalanced case with $\mu_0 \neq 0$, two concentric Fermi surfaces (FSs) generally emerge, which come from the electron and hole pocket (Fig. 4b) respectively. We further consider the inter-layer interaction $H_I = V \sum_{k,p,q,\sigma,\sigma'} c_{+,p,\sigma}^\dagger c_{+,p-q,\sigma} c_{-,k,\sigma'}^\dagger c_{-,k+q,\sigma'}$ on top of $H_0$. At the mean-field level, the phase diagram with varying $\mu_0$ and $E_g$ is shown in Fig. 4e. For the balanced case $\mu_0=0$, the FSs have perfect nesting, leading to the zero-momentum EI, as indicated by Fig. 1a. With increasing $|\mu_0|$ the zero-momentum EI phase is destabilized and FFLO EI is generated [14-21]. With further raising the imbalance, the EIs are finally replaced by the two-carrier region and the electron or hole gas [18].

**The emergent frustration and moat band.** To derive the low-energy effective theory of excitons on moat band beyond mean-field level, we first make Hubbard-Stratonovich decomposition of the electron-hole interaction $H_I$, leading to the functional action as

$$S_I = \int dt \sum_{k,q,\sigma} b_q c_{-,k+q,\sigma}^\dagger c_{+,k,\sigma} + c_{+,k,\sigma}^\dagger c_{-,k+q,\sigma} b_q^\dagger - \frac{1}{V} \int dt \sum_q b_q^\dagger b_q,$$

where $b_q^\dagger$, $b_q$ are the introduced auxiliary bosonic fields. $S_I$ formally describes the fermion-boson interaction. Besides, the action of the free fermions is obtained as

$$S_f = \int dt \sum_{k,q,\sigma} \bar\psi_{k,q,\sigma} G_{f,0}^{-1}(k,q,t) \psi_{k,q,\sigma},$$

where $\bar\psi_{k,q,\sigma}$ and $\psi_{k,q,\sigma}$ are Grassmann spinor fields defined as $\psi_{k,q,\sigma} = [c_{+,k,\sigma}, c_{-,k+q,\sigma}]^T$ and $\bar\psi_{k,q,\sigma} = [\bar c_{+,k,\sigma}, \bar c_{-,k+q,\sigma}]$, and $G_{f,0}(k,q,t)$ is the bare Green's function of the fermions, which is given by $G_{f,0}^{-1}(k,q,t) = (i\partial_t - \xi_{k,\sigma} + \mu_0)(\sigma^z+1)/2 + (i\partial_t + \xi_{k+q,\sigma} + \mu_0)(1-\sigma^z)/2$. Then, we integrate out the fermionic Grassmann fields and make expansion with respect to the fermion-boson coupling, which leads to an effective action of the bosons $S_b$. To the second order, the saddle point equation $\delta S_b/\delta b_q=0$ produces the mean-field equation at the critical point. Interestingly, it is found that the saddle point equation $\delta S_b/\delta b_q=0$ can be simultaneously satisfied by all the momentums on the momentum loop with radius $Q$, i.e., $|q|=Q$, implying that there are an infinite number of degenerate momentum channels for boson condensation. Lifting such degeneracies with condensation on a single momentum point generates the FF state; coherent condensation at two opposite momentum points leads to the LO state. Here, we show that in the intermediate regime between the FFLO and the two-carrier region in the phase diagram of Fig. 4d, correlated physics on the moat band emerges, which can favor topological orders rather than the boson condensations.

To derive the emergent low-energy physics, we further make expansion with respect to the fermion-boson coupling to the fourth order. Besides, since the saddle point equation $\delta S_b/\delta b_q = 0$ is satisfied on the loop $Q$, the momentums of the low-energy excitons lie within a momentum shell around $Q$. Keeping the long-wave modes, one can derive the action describing the fluctuations that are slowly-varying in space and time. The fluctuations in the Gaussian level lead to the first term of Eq. (1), while higher-order fluctuations generate the exciton-exciton interaction, i.e., the second term of Eq. (1). As shown in detail by the Supplementary Information, an effective theory describing interacting excitons on the moat band, *i.e.*, Eq. (1), takes place, which plays the dominant role in the intermediate regime between the FFLO and the two-carrier region.

**The energetics of ETO compared to boson condensation.** When the exciton-exciton interaction $U$ becomes dominant in Eq. (1), a single quantum state could not be occupied by more than one exciton even at zero temperature. In this case, a natural theoretical method for studying possible topological orders is the flux attachment, which represents the bosonic states by spinless fermions coupled to a string operator, namely, $b_r^\dagger = f_r^\dagger e^{iU_r}$ and $b_r = f_r e^{-iU_r}$, where $U_r = \sum_{r'\neq r} \arg(r'-r) f_{r'}^\dagger f_{r'}$, with $\arg(\mathbf{r})$ being the angle of $\mathbf{r}$. $U_\mathbf{r}$ is the string operator that reproduces the bosonic statistics. This is an exact representation that respects the dimension of the local Hilbert space. Accordingly, in the first-quantized form, the exciton wave function is related to the fermion wave function via $\psi_b(r_1,\ldots,r_N) = \prod_{i<j}^N \frac{z_i - z_j}{|z_i - z_j|} \psi_f(r_1,\ldots,r_N)$, where the factor $\prod_{i<j}^N \frac{z_i - z_j}{|z_i - z_j|}$ ensures the flux attachment and introduces the statistical transmutation, and $z_i$ denotes the complex coordinate of the *i*-th particle. The composite fermions could further lower the system energy, since they are free from the energy cost of the exciton-exciton interaction $U$, due to the antisymmetric nature of the fermion wave function.

The attached flux is proportional to the particle density, i.e., $B_{CS} = 2\pi n_b$, and behaves as an emergent uniform magnetic field. Solving the fermion model under the uniform field, the LL can be obtained as $E_l = \frac{Q^2}{2m_b}\left\{\left[\frac{(l+\frac{1}{2})\omega_c}{\frac{Q^2}{2m_b}}\right]^{\frac{1}{2}} - 1\right\}^2 - \mu_b$, where is $\omega_c$ the cyclotron frequency, and the eigenstates are given by the determinant, $\psi_f^l(r_1,\ldots,r_N) = \det_{m,j}[\chi_m^l(z_j)]/\sqrt{N!}$, where $\chi_m^l(z_j) \propto L_l^m\left(\frac{|z|^2}{2l_B^2}\right)$. $L_l^m(x)$ is the adjoint Laguerre polynomial and $l_B = 1/\sqrt{2\pi n_b}$. Then, the ETO energy can be evaluated as $E_{\text{ETO}} = \langle \psi_b | H_b | \psi_b \rangle = \left(\frac{\pi^2 n_b^2}{2m_b Q^2}\right)\log^2\left(\frac{4n_b}{Q^2}\right)$. This agrees with the result obtained in [10] for the topologically ordered state of bosons. We recall that the FF or LO state has the energy $E_{\text{FFLO}} \propto \frac{U n_b}{m_b}$ [8,14,15], and the non-uniform condensation state possesses the energy $E_{\text{NU}} \propto \frac{U n_b^{4/3}}{m_b Q^{2/3}}$ [40]. In comparison, the ETO has the lowest energy under relatively low density, i.e., the green area in Fig. 4d.

**Explanation of experiments using the ETO theory.** The damping term implicit in Eq. (1) brings about the finite lifetime *τ*. Within *τ*, the coherence is preserved in the edge channels, which indicates the gapless nature of the edge state. Then, the recombination of electrons and holes results in virtual contacts [42] with the spacing $\lambda = v_F \tau$, where $v_F$ is the edge Fermi velocity. This leads to the non-quantized resistance in macroscopic devices as shown in Extended Data Fig. 3. Interestingly, since temperatures below the gap barely affect the recombination rate, $\lambda$ should be insensitive to temperatures, which is also observed, as shown in Extended Data Fig. 3. In addition, under more negative $V_f$, the recombination is easier to occur between the two layers, which enhances backscattering and could explain the tilted slope of $R_{xx}^{EI}$ in Fig. 2d. The ETO theory can also explain the detailed features of the measured gap energies $\Delta$ in Fig. 2a. $\Delta$ reflects the energy scale that excites charged quasi-particles and quasi-holes from the ETO, which is proportional to the emergent flux $B_{CS}$. Because the imbalance suppresses the formation of e-h pairs, $B_{CS}$ decreases with increasing the imbalance, leading to the decaying behavior of $\Delta$ as $V_f$ is being tuned away from the CNP (see Sec. V of Supplementary Information). Moreover, an important feature here is that the ETO bulk gap is independent of the voltage range since the latter is only determined by the critical imbalance within which the ETO is favored. This can explain the different $B_\perp$-dependence of the EI gap and the voltage range shown in Extended Data Fig. 9.

**Calculations of ETO edge transport based on the LB formula**. The edge states of ETO can be spatially separated under $B_\perp$. The separation driven by the Lorentz force is in the xy-plane, as shown by the inset to Extended Data Fig. 6a. In comparison, the weak pairing between the electron and hole edge channel is mainly along the z-direction, *i.e.*, perpendicular to the layers. To estimate the edge separation under $B_\perp$, we start with a semiclassical analysis. For $B_\perp = 0$, the two edge channels overlap in real space, and both of them display a characteristic localization length from the boundary, $r_0$. In the semiclassical picture, the two edges are formed by the repeated reflections of the cyclotron motion upon the boundary, driven by the emergent gauge field $B_{CS}$. We can regard the electrons and holes as being effectively attached to opposite Chern-Simons fields $B_{cs}^e \approx -B_{cs}^h \approx \pi m v_F/er_0$ where $v_F$ is the Fermi velocity. Then, by applying external field $B_\perp$, the total field seen by the electron and hole channel become different, namely, $B^h = \pi m v_F/er_0 + B_\perp$ and $B^e = -\pi m v_F/er_0 + B_\perp$. This results in different cyclotron radii for the two-channel modes, and the separation is obtained as

$$d(B_\perp) = \left| \frac{\pi m v_F}{-eB_\perp + \pi m v_F/r_0} - \frac{\pi m v_F}{eB_\perp + \pi m v_F/r_0} \right|.$$

In the following, we use parameters extracted from experiments. The Fermi velocity $v_F = 2\times 10^4$ m/s. The decay length of the edge state into the bulk is taken as $r_0 = 10$ nm. The effective mass of edge Dirac particles is defined by $m = \hbar k_F/v_F$, with $k_F$ the Fermi wave number, which leads to $m \approx 0.46 m_0$ with $m_0$ the bare electron mass.

On the other hand, the ballistic transport of edge states is described by LB formula that defines electron transport in multi-terminal devices. The current and the probe voltages satisfy:

$$I_{qp} = \frac{e^2}{\hbar} \sum_q T_{qp} V_p - T_{pq} V_q,$$

where $q$ and $p$ are contact labels, and $T_{pq}$ is the transmission probability from contact $p$ to $q$. If the transmission from one contact $p$ to the next one $q$ is perfect, we have $T_{pq} = 1$. The total current is conserved so that $\Sigma_q I_{pq} = 0$. We then consider a six-terminal Hall bar device with virtual contacts (Sec. II of Supplementary Information). The number of virtual contacts is estimated from the actual sample size compared to the coherent length $\lambda$. Note that two channels are forming effective helical transport. The transmission rate between the outer loop and the contact is given by $T_{n+1,n} = 1$, while that between the inner loop and the contact is given by $T_{n+1,n} = x$, where $x$ is generally less than 1 because it decays with the channel separation induced by $B_\perp$. Besides, the transmission rate between the channels and the virtual contacts is assigned to be $y$. The exponential decay of the transmission is assumed, *i.e.*, $x = y = e^{-d(B_\perp)/\xi}$, where $\xi$ is the characteristic decay length.

Using the LB formula, we numerically calculate $R_{xy}$ and $R_{xx}$ as a function of $B_\perp$ for the separated edges. The characteristic length of the transmission is used as $\xi \sim 80$ nm. Extended Data Fig. 6 displays a good agreement between the calculations and the data extracted from Fig. 2b and that from the Hall bar device C in Extended Data Fig. 4. For strong $B_\perp$, we find that the separation can reach up to a few micrometers, generating a chiral-like edge transport (Sec. II of Supplementary Information). This well accounts for $\sigma_{xy} \sim h/e^2$ observed in the EI regime in Figs. 3b and 3c. Thus, the observed helical-like to chiral-like edge transport evolution under rising $B_\perp$ reveals the key features of the ETO edge state.

**Method references**

**Acknowledgments**

This work at Nanjing University was supported by the National Key R&D Program of China (Grant No. 2022YFA1403601) and the National Natural Science Foundation of China (Grant No. 12274206, No. 12074177, No. 12034014), Program for Innovative Talents and Entrepreneur in Jiangsu and the Xiaomi Foundation. The work at Peking University was supported by the Strategic Priority Research Program of the Chinese Academy of Sciences (Grant No. XDB28000000), National Key R&D Program of China (Grant No. 2019YFA0308400), and National Natural Science Foundation of China (Grant No. 11921005). The InAs/GaSb quantum wells structures were prepared by molecular beam epitaxy by Gerard Sullivan. A portion of this work was performed at the National High Magnetic Field Laboratory, which is supported by the National Science Foundation Cooperative Agreement No. DMR-1157490 and the State of Florida.

**Author contributions**

R. W., T. A. S., and B. W. conceived the theoretical project. R. W. developed the theoretical model and performed the calculations supervised by B. W. R.-R. D. and L.-J. D. conceived the experimental project. L.-J. D. fabricated devices and performed transport experiments. L.-J. D. and R.-R. D analyzed the data. R. W., L.- J. D., and R.-R. D. co-wrote the manuscript with input from other authors. All authors discussed the results. R.-R. D. provided overall coordination of the whole project.

**Competing interests**
The authors declare no competing financial interests.

**Corresponding authors**
Correspondence to Baigeng Wang, Lingjie Du, Rui-Rui Du

**Additional information**

Supplementary information is available in the online version of the paper.
Reprints and permission information are available online.

**Data availability**

All data needed to evaluate the conclusions in the paper are included in this paper. Additional data that support the plots and other analyses in this work are available from the corresponding authors upon request.

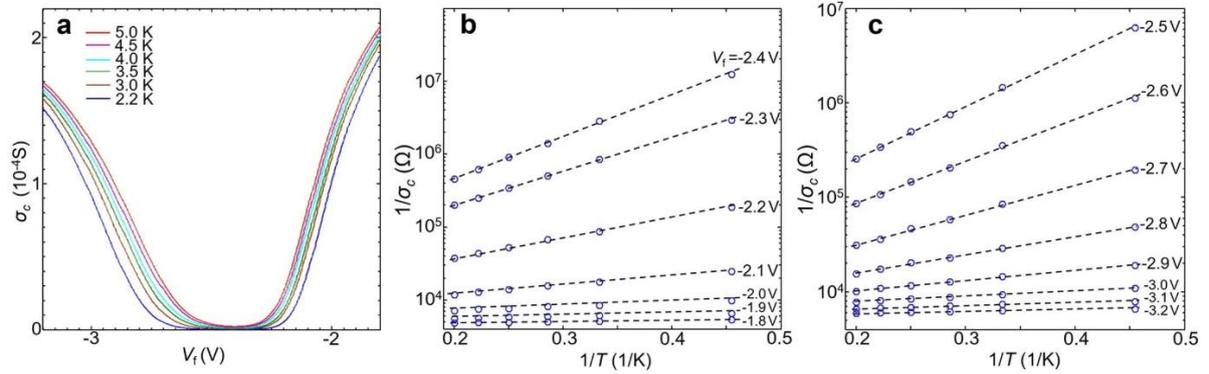

Extended Data Fig.1 **Extraction of energy gap $\Delta$ under 0 T.** (a). Bulk conductance of the Corbino device $\sigma_c$ as a function of $V_f$ under 0 T at different temperatures. (b) and (c). Arrhenius plot of the conductance $\sigma_c$ at $V_f$ from -1.8 V to -2.4 V, and from -2.5 V to -3.2 V, respectively. The data can be fit by $\sigma_c \propto \exp(-\Delta/2k_bT)$ to obtain $\Delta$. Dashed lines are guides to the eye. For $V_f \geqslant -2.1$ V or $V_f \leqslant -2.9$ V, the $\sigma_c$ exhibits non-activated temperature dependence, which is also plotted in panel (b) for comparison. Note that according to the ETO theory, as the e-h density imbalance increases, the gap that opens around the Fermi level would decrease towards zero, which could describe the data in $V_f \geqslant -2.1$ V or $V_f \leqslant -2.9$ V.

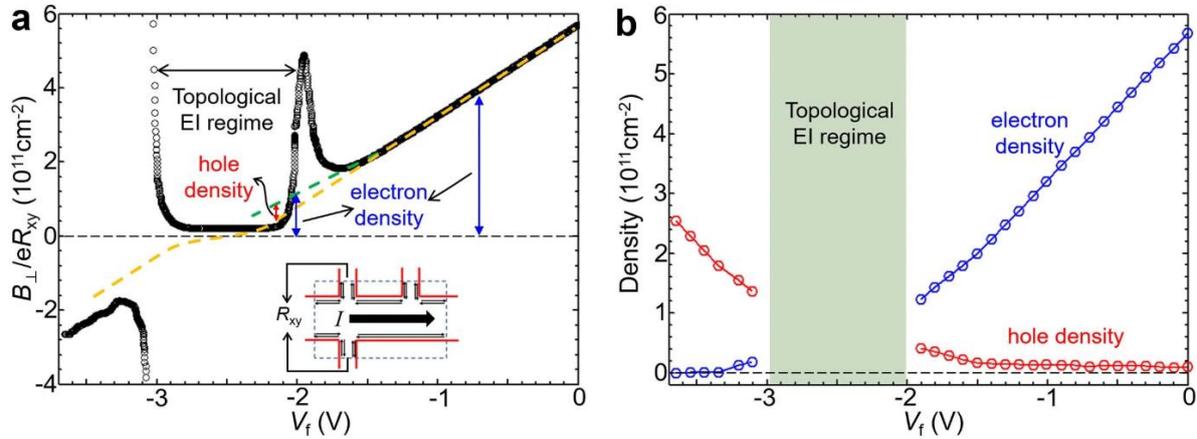

Extended Data Fig.2 **Estimation of carrier densities.** (a). $B_\perp/eR_{xy}$ versus $V_f$ in a 50 μm × 50 μm Hall-bar at 300 mK under $B_\perp = 1$ T. Inset shows a schematic of the asymmetric Hall-bar, and the dashed box region is covered by the front gate. The yellow dashed line is obtained from the integration of measured capacitance over $V_f$, representing the net-carrier density $|n_e-n_p|$ (see Ref. 12). In the linear regime of $B_\perp/eR_{xy}$, $n_p$ is negligible with $|n_e-n_p| \approx n_e$, and the yellow line represents $n_e$ (blue arrow). $B_\perp/eR_{xy}$ deviates from the yellow line due to the increased $n_p$ in the two-carrier transport. In this regime, $V_f$ would tune both $n_e$ and $n_p$. Since the geometry capacitance remains constant and dominates the measured capacitance, $|n_e-n_p|$ would still change linearly with $V_f$ and follow the yellow line. The green dashed line represents the $V_f$-dependence of $n_e$ (blue arrow) in the two-carrier transport regime [12]. The difference between the green and yellow lines corresponds to $n_p$ (red arrow). At $V_f = -2$ V, $B_\perp/eR_{xy}$ drops because the topological edge occurs and contributes a large $R_{xx}$ component on the $R_{xy}$ due to the asymmetric Hall bar geometry [12]. The edge state persists in a voltage range from -2 V to -3 V (topological EI

regime). At $V_f$ = -2 V, $n_e \approx 1.1 \times 10^{11}$ cm$^{-2}$ and $n_p \approx 4.5 \times 10^{10}$ cm$^{-2}$. From -2 V to -2.5 V, the e-h density imbalance regime with electrons as the majority carrier overlaps the topological EI regime. Symmetrically, a similar case occurs from -2.5 V to -3 V but with holes as the major carrier. (b). Calculated $n_e$ and $n_p$ using a two-carrier model (see Methods). The calculated results agree well with those obtained in (a).

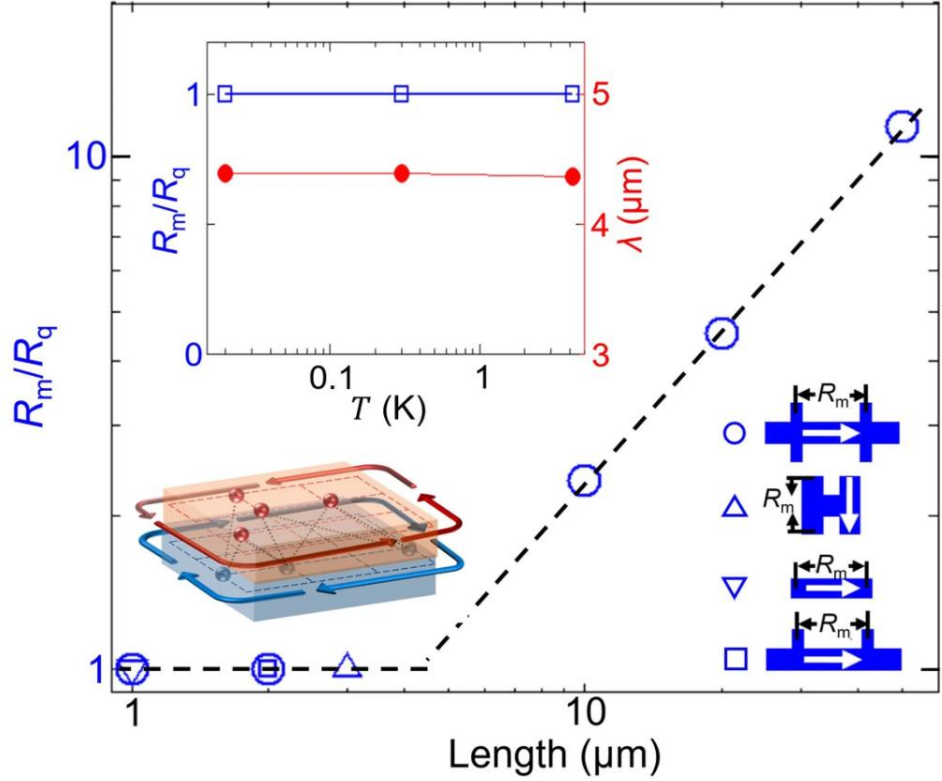

Extended Data Fig.3 **Summary of edge resistance against edge length measured in the topological EI from shallowly-inverted InAs/GaSb QWs with different device geometries.** $R_m$ is the measured resistance. $R_q$ denotes quantized values calculated by the LB formula, *i.e.*, $h/2e^2$ (Hall bar, circle) [23], $h/4e^2$ (H bar, up-triangle), $h/2e^2$ (two terminal bar, down-triangle) [23], and $h/4e^2$ (π bar, square) [23]. The edge lengths are depicted between black arrows in the devices. White arrows show the direction of the current. The transport of the edge state in the mesoscopic devices with length <λ is quantized. In the macroscopic devices, the backscattering occurs between two channels through virtual contacts [42] where channels equilibrate over λ; within λ, the helical-like edge transport is ballistic and gives the value of a quantum resistor $h/2e^2$. The up-left inset shows that the quantized resistance in the π bar device and the extracted coherence length λ from the macroscopic Hall bars are independent of temperatures up to 4 K. The lower-left inset shows the schematic plot of counter-propagating edge channels with a bulk EI state and arrows on the perimeters indicate the helical-like feature of the edge channels.

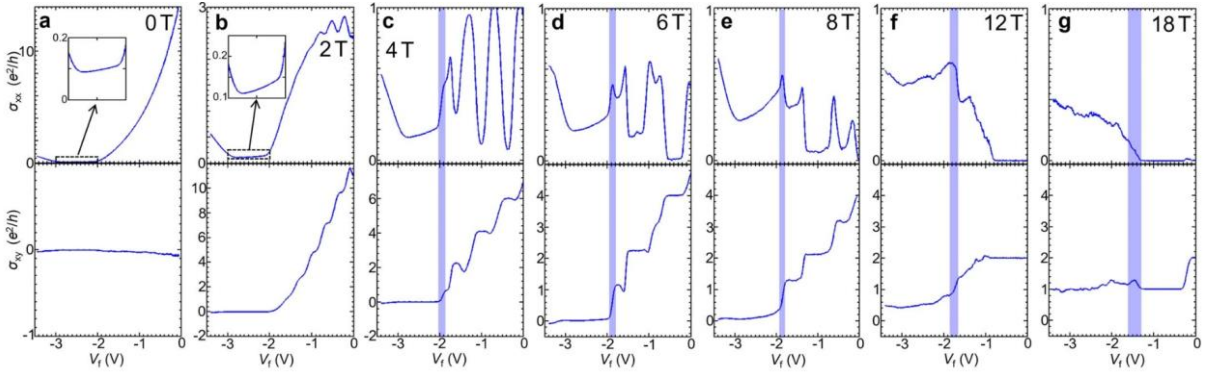

Extended Data Fig.4 **Longitudinal (top panel) and Hall (bottom panel) conductivities under magnetic fields up to 18 T.** (a-g) Longitudinal (top panel) and Hall (bottom panel) conductivities, $\sigma_{xx}$ and $\sigma_{xy}$, converted from $R_{xx}$, $R_{xy}$ measured in a 50 μm x 50 μm Hall-bar device C as a function of $V_f$ at 20 mK under $B_\perp$ up to 18 T. Magneto-transport below 8 T repeat that in device A (Fig. 2). The purple marks the transition regime from the QH states to topological EI. At 4 T, a $\sigma_{xx}$ peak starts to emerge in the transition regime, and $\sigma_{xy}$ collapses. The $\sigma_{xx}$ peaks become more visible for higher $B_\perp$. Meanwhile, $\sigma_{xy}$ in the topological EI regime increases because the edge channels start to separate under $B_\perp$. At 18 T, $\sigma_{xy}$ between the $v = 1$ QH state and topological EI is close and the $\sigma_{xy}$ collapse is replaced by a bump due to two-carrier transport. The corresponding $\sigma_{xx}$ peak is smeared due to the residual conductance in the topological EI regime. $\sigma_{xx}^{EI}$ (in the topological EI regime) starts to decrease at 18 T and approaches zero at 35 T (Fig. 3c). The shrinking of the inner loop has two results: the inner loop is separated from the outer loop (at low $B_\perp$); the inner loop is separated from the contacts and the outer loop (at high $B_\perp$). The two results have opposite effects on $\sigma_{xx}^{EI}$ in a macroscopic Hall-bar ($\sigma_{xx}^{EI} < e^2/2h$ at 0 T): the former makes $\sigma_{xx}^{EI}$ increase towards $e^2/2h$ due to reduced backscattering between the edge channels; the latter makes $\sigma_{xx}^{EI}$ decrease to zero since backscattering between the inner loop and the contacts decreases, and the outer loop works like a single chiral edge.

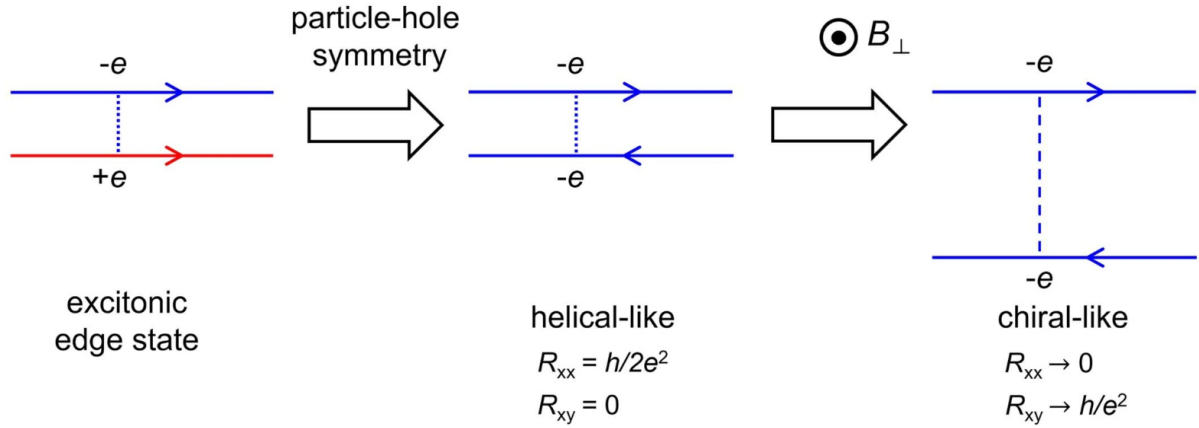

Extended Data Fig.5 **Schematic plot illustrating the ETO edge state on one side of a Hall bar.** Left and Middle, $B_\perp = 0$: The bulk of the imbalanced electron-hole bilayer opens a many-body gap, with its edge state formed by a pair of the weakly-bound particle (-$e$) and hole (+$e$) channels, as demonstrated by the theory of ETO; the robustness of the edge states is protected by a BCS-like bulk gap. Considering the quasiparticle-quasihole symmetry near the Fermi level, the excitonic edge state is viewed as equivalent to a helical-like state containing two oppositely propagating (-$e$) channels (Sec. V of Supplementary Information). Right, when $B_\perp$ is applied, the two channels are spatially separated due to the Lorentz force, forming inner- and outer- loops in a Hall bar. The dotted lines denote the scattering between the two channels, which decays with increasing $B_\perp$ but always exists in realistic samples, explaining why the Hall resistance does not exactly quantize in chiral-like transport.

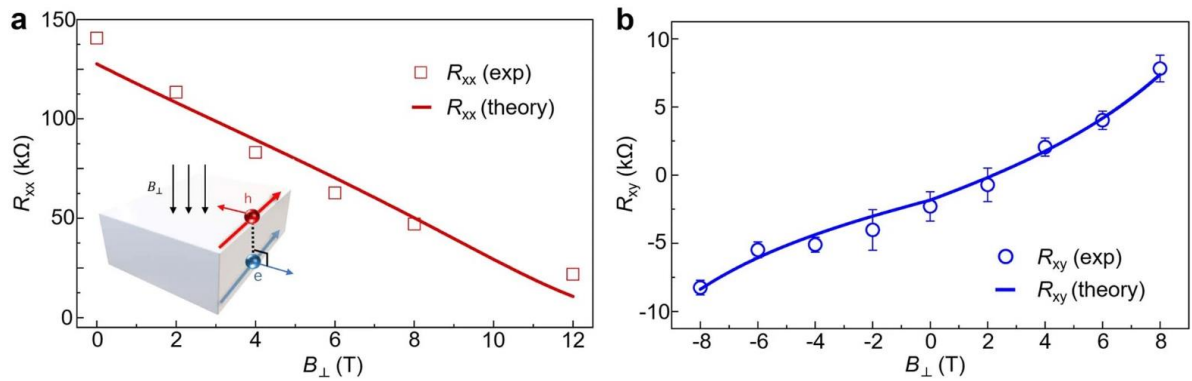

Extended Data Fig.6 **Calculation of the ETO edge transport under $B_\perp$.** To calculate the edge resistance, we consider a six-terminal Hall bar with edge separation under $B_\perp$ and virtual contacts (See Methods). (a) and (b) respectively show the calculated $R_{xx}$ and $R_{xy}$ as functions of $B_\perp$ compared with the experimental data. Data $R_{xx}$ in (a) is measured from the Hall-bar device C (see Extended Data Fig. 4), and data in (b) is extracted from $R_{xy}$ in Fig. 2b. The inset to (a) schematically demonstrates the effect of $B_\perp$ on the two edge channels. The error bars in (b) come from the uncertainties in the experimental $R_{xy}$ in the topological EI regime.

| | Electronic FQH | Excitonic FQH (ETO) |
|---|---|---|
| Constituents | Electron | Exciton |
| Origin of Landau quantization | External field $B_\perp$ | "moat" + correlation $U$ → Landau level at zero field $B_\perp = 0$ |
| Bulk Low-energy excitation | Laughlin liquid (e.g. $m = 3$)<br>Charge: $Q = e/3$<br>Statistics: anyons with $\phi = \pi/3$ | Excitonic Laughlin liquid ($m = 2$)<br>Charge: $Q = 0$<br>Statistics: anyons with $\phi = \pi/2$ |
| Chern-Simons transmutation | Electron bound to 2-flux quanta; QH of CFs ($\nu = 1$) | Exciton bound to 1-flux quanta; Quantum anomalous Hall of CFs ($C = 1$) |
| Edge excitation and edge state conductance | Chiral mode of CFs:<br>$R_{xx} = 0$<br>$R_{xy} = 3h/e^2$ | Manifest as helical edge in conductance measurement ($B_\perp = 0$)<br>$R_{xx} = h/2e^2$ (for small Hall bar)<br>$R_{xy} = 0$ |

Extended Data Fig.7 **Summary of the ETO properties.** We make a conceptual analogy of ETO with the electronic $1/m$ Laughlin liquid that is formed in an external magnetic field.

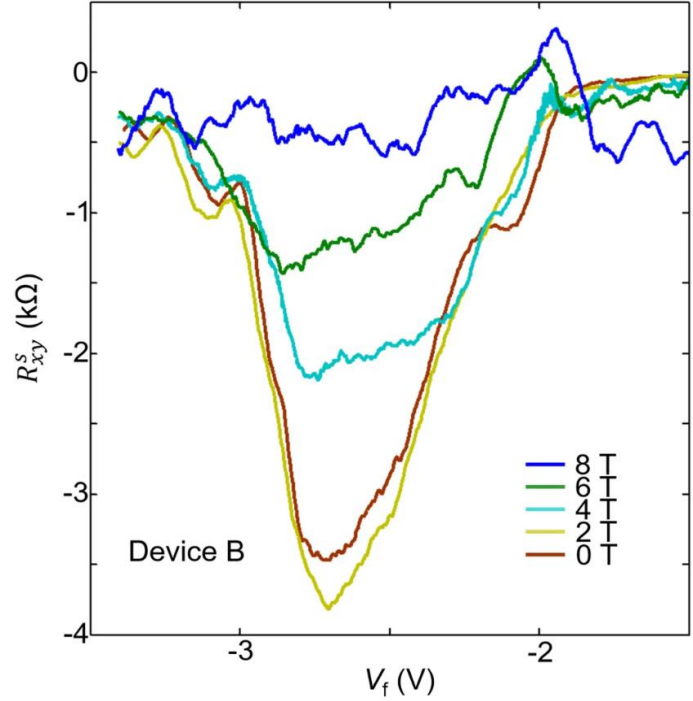

Extended Data Fig.8 **The symmetric term $R^s_{xy}$ from $R^{EI}_{xy}$.** $R^s_{xy}$ is shown as a function of $V_f$ at different $B_\perp$ in device B.

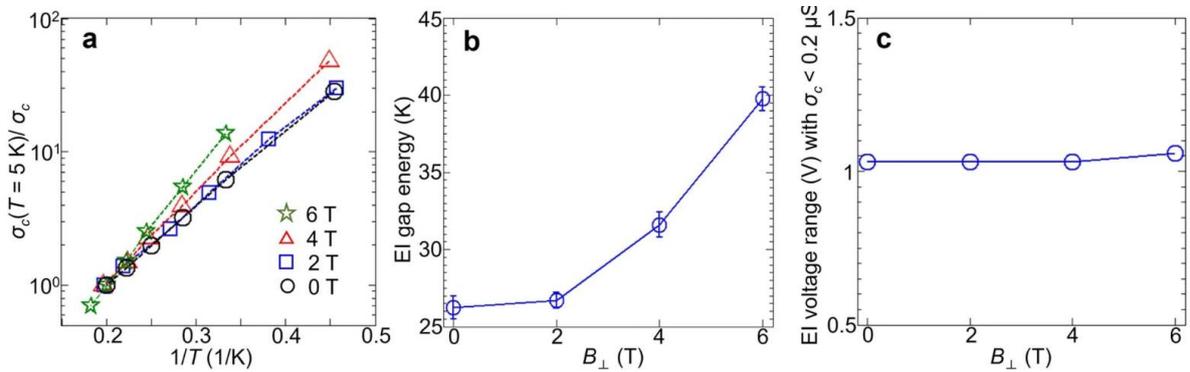

Extended Data Fig.9 **Energy gap under different perpendicular magnetic fields $B_\perp$.** (a) The Arrhenius plot of the Corbino conductance $\sigma_c$ at the CNP with temperatures under fixed $B_\perp$. The Corbino conductance is normalized to the values at 5 K. The EI gap energy under a magnetic field is deduced through $\sigma_c \propto \exp(-\Delta/2k_bT)$, their values are displayed in (b). The error bars come from the uncertainty in the extraction of gap energy from the Arrhenius plot. (c) shows the voltage ranges where the state in the topological-EI regime is insulating ($\sigma_c < 0.2\mu S$) under $B_\perp$. Obviously, as $B_\perp$ increases, the EI gap becomes larger, but the voltage ranges of the insulating state in the topological-EI regime remain nearly the same.